\documentclass[12pt]{article}

\usepackage{scicite}

\usepackage{times}
\usepackage{amsmath,epsfig,wrapfig,graphicx,color,bm}

\usepackage[dvipdfmx]{xcolor}

\newenvironment{sciabstract}{%
\begin{quote} \bf}
{\end{quote}}

\title{Novel Excitations near Quantum Criticality in Geometrically Frustrated Antiferromagnet CsFeCl$_{3}$\\
{\small [{\bf One sentence summary}] Nontrivial hybridization of longitudinal and transverse spin fluctuations induced by noncollinearity of a spin structure is reported.} }

\author
{Shohei Hayashida,$^{1}$ Masashige Matsumoto,$^{2}$ Masato Hagihala,$^{1}$ \\
 Nobuyuki Kurita,$^{3}$ Hidekazu Tanaka,$^{3}$  
Shinichi Itoh,$^{4}$ Tao Hong,$^{5}$ Minoru Soda,$^{1}$ \\
Yoshiya Uwatoko,$^{1}$ and Takatsugu Masuda$^{1\ast}$\\
\\
\normalsize{$^{1}$Institute for Solid State Physics, The University of Tokyo, Chiba 277-8581, Japan}\\
\normalsize{$^{2}$Department of Physics, Shizuoka University, Shizuoka 422-8529, Japan}\\
\normalsize{ $^{3}$Department of Physics, Tokyo Institute of Technology, Meguro-ku, Tokyo
 152-8551, Japan}\\
\normalsize{ $^{4}$Neutron Science Division, Institute of Materials Structure Science,}\\
\normalsize{High Energy Accelerator Research Organization, Tsukuba, Ibraki 305-0801, Japan}\\
\normalsize{$^{5}$Neutron Scattering Division, Oak Ridge National Laboratory, 
Oak Ridge, Tennessee 37831, USA}\\
\\
\normalsize{$^\ast$To whom correspondence should be addressed; E-mail: masuda@issp.u-tokyo.ac.jp.}
}

\date{}

\begin{document} 


\baselineskip24pt


\maketitle 

\renewcommand{\H}{{\mathcal H}}

\newcommand{\bS}{{\bm{S}}}
\newcommand{\bW}{{\bm{W}}}

\newcommand{\bk}{{\bm{k}}}
\newcommand{\br}{{\bm{r}}}
\newcommand{\bd}{{\bm{d}}}
\newcommand{\bq}{{\bm{q}}}
\newcommand{\bQ}{{\bm{Q}}}

\newcommand{\ba}{{\bm{a}}}
\newcommand{\bb}{{\bm{b}}}
\newcommand{\bc}{{\bm{c}}}
\newcommand{\bA}{{\bm{A}}}
\newcommand{\bB}{{\bm{B}}}
\newcommand{\bu}{{\bm{u}}}
\newcommand{\bv}{{\bm{v}}}
\newcommand{\balpha}{{\bm{\alpha}}}
\newcommand{\bU}{{\bm{U}}}
\newcommand{\bV}{{\bm{V}}}

\newcommand{\hq}{{\hat{q}}}

\newcommand{\TN}{T_{\rm N}}

\newcommand{\Sr}{{Sr$_2$CoSi$_2$O$_7$}}
\newcommand{\Ba}{{Ba$_2$CoGe$_2$O$_7$}}
\newcommand{\Tl}{{TlCuCl$_3$}}
\newcommand{\Cs}{{CsFeCl$_3$}}
\newcommand{\Rb}{{RbFeCl$_3$}}
\newcommand{\BaCo}{{Ba$_3$CoSb$_2$O$_9$}}


\begin{sciabstract}
Investigation of materials that exhibit quantum phase transition provides valuable insights into fundamental problems in physics. We present neutron scattering under pressure in a triangular-lattice antiferromagnet which has a quantum disorder in the low-pressure phase and a noncollinear structure in the high-pressure phase. The neutron spectrum continuously evolves through the critical pressure; a single mode in the disordered state becomes soft with the pressure, and it splits into gapless and gapped modes in the ordered phase. Extended spin-wave theory reveals that the longitudinal and transverse fluctuations of spins are hybridized in the modes because of the noncollinearity, and novel magnetic excitations are formed. We report a new hybridization of the phase and amplitude fluctuations of the order parameter in a spontaneously symmetry-broken state.
\end{sciabstract}


For the understanding of condensed matter, investigation of the collective excitation in low energy 
range is indispensable. According to the quantum field theory, the excitation in the system with 
spontaneously symmetry broken is characterized by the phase and amplitude fluctuations of the order parameters. 
The former is known as the Nambu-Goldstone (NG) mode, and the latter is called as the amplitude mode. 
Even though these modes are usually separated, they are hybridized under some conditions, 
and interesting phenomena are induced; for example in a crystal lattice system, acoustic phonon (NG mode) and optical phonon (amplitude mode) are hybridized thorough anharmonic terms in a thermoelectric material PbTe, 
which renormalizes the phonon spectrum and leads to low thermal 
conductivity and high figure of merit in thermoelectric property~\cite{Delaire11}. 
Such a hybridization effect could exist in other types of elementary 
excitations, but no research has been reported to our knowledge. 
One of the reasons is that the amplitude mode itself is not trivial in other systems: 
superconductors~\cite{Matsunaga14}, charge density wave~\cite{Pouget91}, 
ultra-cold atoms~\cite{Endres12}, and insulating antiferromagnets. 
The existence of the amplitude mode in the antiferromagnets requires strong 
fluctuation of magnetic moment and the system location near a quantum critical point (QCP). 
So far it has been found in a number of spin systems free from geometrical frustration 
such as quasi-one-dimensional chains~\cite{Lake00,Zheludev02}, dimer~\cite{Ruegg2004,Ruegg2008,Merchant2014}, square lattice~\cite{Jain2017}, and two-leg ladder antiferromagnet~\cite{Hong2017}. 
In magnets in the presence of geometrical frustration which could induce the hybridization of the modes, 
on the other hand, magnon excitations from ordered states have been less focused these days~\cite{Ito2017}, 
even though fractional excitations in spin liquid have been intensively studied~\cite{Han2012,Fak2012,Paddison2017}. Particularly, the collective excitations from a noncollinear spin structure in geometrical frustrated lattice near QCP has not been studied, and its investigation is of primary importance to discover a novel state and to advance the physics of the frustration and quantum criticality.

\begin{figure}[tbp]
\begin{center}
\epsfig{file=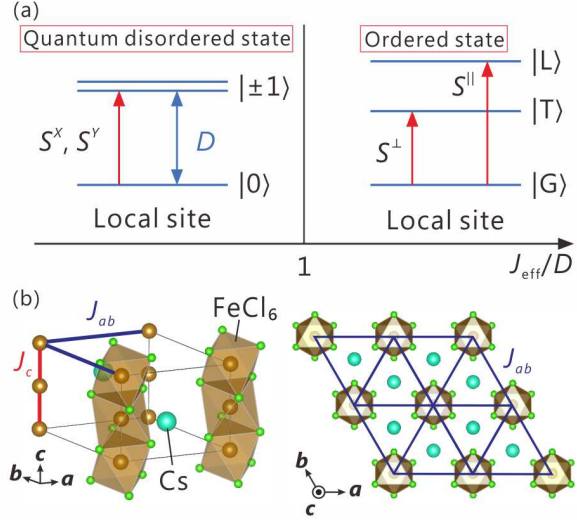,width=0.5\hsize}
\caption{(a) Schematic diagram of the $S=1$ easy-plane antiferromagnet. 
In the ordered state, the doublet excited states $|\pm 1 \rangle$ splits into $|{\rm L} \rangle$ and $|{\rm T} \rangle$.
Here, the former and latter have longitudinal and transverse fluctuations, respectively.
(b) Crystal structure of CsFeCl$_3$ with the space group $P6_{3}/mmc$~\cite{Kohne1993}. 
Magnetic Fe$^{2+}$ ions having pseudo-spin $S=1$ form one-dimensional 
chains along the crystallographic $c$ axis, and the chains form the triangular lattice in the 
$ab$ plane. 
Red and blue lines indicate the intrachain interaction $J_c$ and 
interchain/intratriangle interaction $J_{ab}$, respectively.}
\label{fig1}
\end{center}
\end{figure}

Spin $S=1$ easy-plane antiferromagnet is one of the prototypical quantum spin systems 
that allow to explore nature of the quantum phase transition (QPT)~\cite{Matsumoto2007,Matsumoto2014}.
The Hamiltonian is expressed by 
$
{\cal H}=\sum_{i}D(S_{i}^{Z})^{2}+\sum_{{i,j}}J_{ij}{\bm S}_{i}\cdot{\bm S}_{j},
\label{Hamiltonian}
$
where ${\bm S}_{i}$ is the spin operator at the $i$ site.
The first term $D$ is positive and gives an easy-plane single-ion anisotropy.
The second term is a Heisenberg interaction with $J_{ij}>0$ for an antiferromagnetic coupling.
The positive $D$ splits the triplet $S$ = 1 states into the singlet ground state $S^{Z}=0$ 
and the doublet excited states $S^{Z}=\pm 1$, 
and favors a quantum disordered (QD) state as shown in the left in Fig.~\ref{fig1}(a).
By contrast, the spin interaction $J_{ij}$ allows the system to occur the magnetic long-range order (LRO).

In the LRO phase, the energy eigenstates changes as shown in the right in Fig.~\ref{fig1}(a).
Here, $|G\rangle$ is the ground state, while $|T\rangle$ and $|L\rangle$ are excited states.
They are given by
$|G\rangle = u|0\rangle + v/\sqrt{2}( |1\rangle +  |-1\rangle)$,
$|T\rangle = 1/\sqrt{2}(-|1\rangle +  |-1\rangle)$,
and
$ |L\rangle = -v  |0\rangle + u/\sqrt{2}( |1\rangle +  |-1\rangle)$
with $u^2+v^2=1$.
$u$ and $v$ are determined by $D$ and $J_{ij}$.
In the ordered phase, $v\neq 0$ and a finite magnetic moment appears in the $ab$ plane~\cite{Matsumoto2007,Matsumoto2014}.
We separate the spin operator into longitudinal ($S^\parallel$) and transverse ($S^\perp$) components relative to the local ordered moment.
Remarkable feature is that the second excited state  $|L\rangle$ can be excited only by the longitudinal component $S^\parallel$,
while the first one $|T\rangle$ can be excited by the transverse component $S^\perp$.

The excited states at a local site are, thus, separated into two states having longitudinal and transverse fluctuations. 
Indeed the studies in the square lattice predicted an enhanced amplitude/longitudinal mode as one-magnon excitation in 
addition to the NG/transverse mode in the LRO phase near the QCP~\cite{Matsumoto2007,Jain2017}. 
In the collinear magnetic structure in general, the transverse and longitudinal fluctuations are not hybridized,
and they are separated~\cite{Lake00,Zheludev02,Ruegg2004,Ruegg2008,Merchant2014,Jain2017,Hong2017}. 
In the geometrically frustrated spin system, on the other hand,
these fluctuations could be hybridized because of the noncollinearity of the magnetic structure.
This point was theoretically investigated in the continuum-like spectra in the $S=1/2$ and $S=3/2$ systems,
where the longitudinal fluctuation stems from two-magnon process~\cite{Mourigal13}.
Since $S=1$ easy-plane antiferromagnet has the longitudinal fluctuation in one-magnon process, in contrast,
the two fluctuations hybridize in one-magnon level and a novel excited state may appear as a well-defined eigen mode.
Detail of such a study, however, has not been reported neither in experiment nor theory.

CsFeCl$_3$ is  a model material for the $S=1$ easy-plane triangular antiferromagnet as shown in Fig.~\ref{fig1}(b).
The inelastic neutron scattering (INS) study at ambient pressure revealed that the ferromagnetic chains
along the $c$ axis are antiferromagnetically coupled in the $ab$ plane~\cite{Yoshizawa1980}.
The ground state is QD state because of large single-ion anisotropy.
The magnetic susceptibility measurement under pressures exhibited
a pressure-induced magnetic order above a critical pressure of about 0.9 GPa~\cite{Kurita2016}.
Owing to the strong easy-plane anisotropy, the ordered moment aligns in the $ab$ plane.
The neutron diffraction evidenced the noncollinear 120$^{\circ}$ structure in the LRO phase~\cite{Hayashida2018}.
CsFeCl$_{3}$ is, thus, a promising host for the pressure-induced QPT in the geometrically frustrated lattice.

\begin{figure*}[hbtp]
\begin{center}
\epsfig{file=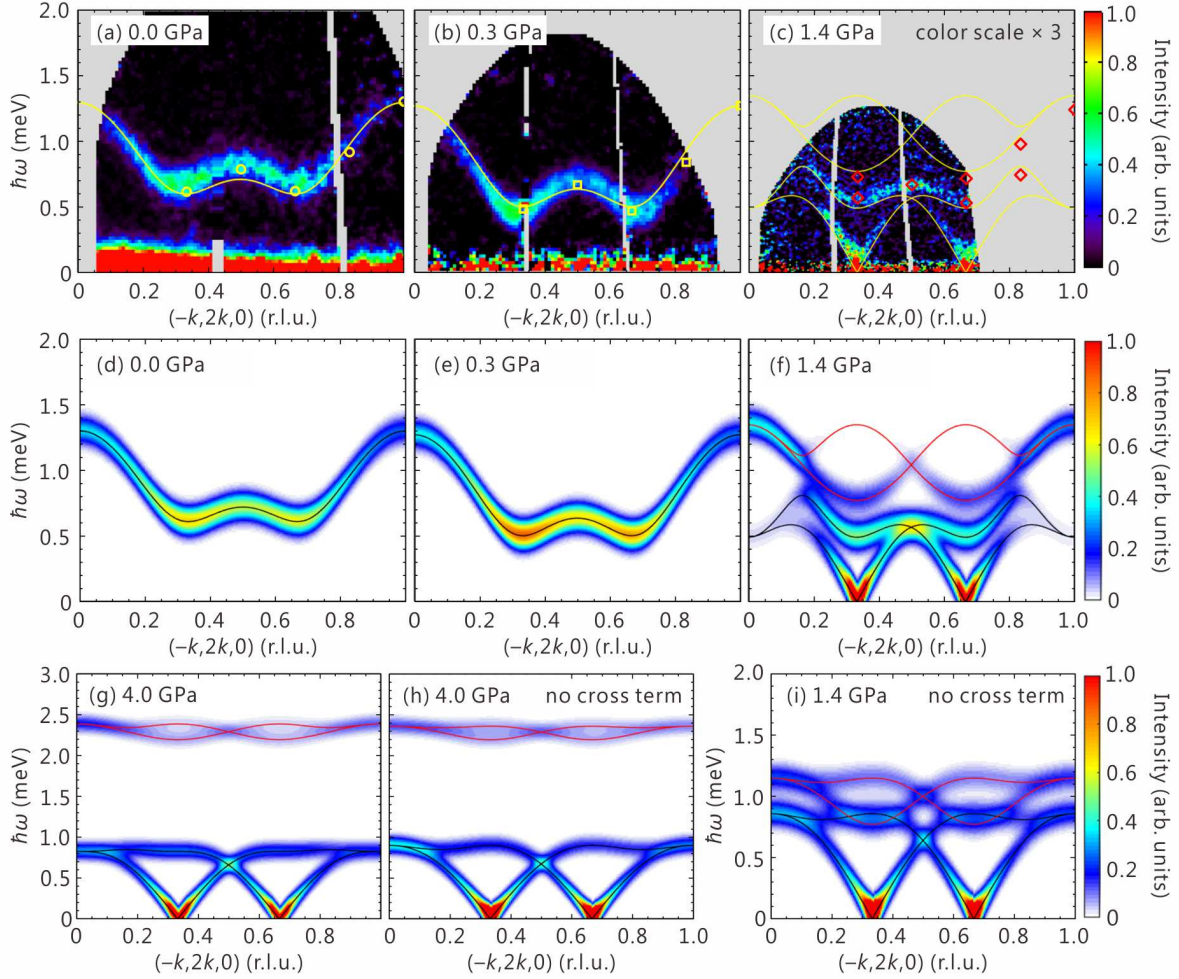,width=1.0\hsize}
\caption{Inelastic neutron scattering spectra obtained at a chopper spectrometer 
under (a) 0.0 GPa at 6 K, 
(b) 0.3 GPa at 2.7 K and (c) 1.4 GPa at 0.9 K 
sliced by the energy transfer - wave vector ($\hbar\omega$ - $\bm q$)  
plane for ${\bm q} = (-k,~2k,~0)$. 
The yellow circles, squares, and red diamonds are the peak positions of 
the excitations obtained from the constant-${\bm q}$ scans using a 
triple-axis spectrometer. 
The solid yellow curves are the dispersions calculated by ESW. 
Calculated neutron cross-section by the ESW under (d) 0.0 GPa, (e) 0.3 GPa, (f) 1.4 GPa, 
and (g) 4.0 GPa at 0 K. 
Calculated neutron cross-section in the absence of the cross term in Eq. (2)
under (h) 4.0 GPa and (i) 1.4 GPa at 0 K.
The black and red solid curves in (f)-(i) are gapless and gapped modes, respectively. 
More detailed pressure dependence of the calculated spectra is shown in Fig. S3 in 
supplementary material. 
}
\label{fig2}
\end{center}
\end{figure*}


The INS spectrum measured at 0.0 GPa by using a chopper spectrometer 
in Fig.~\ref{fig2}(a) exhibits a single dispersive excitation with 
the energy gap of 0.6 meV at the wave vectors ${\bm q} = (-k, 2k, 0)$ for 
$k = 1/3$ and 2/3, which is consistent with previous report~\cite{Yoshizawa1980}. 
The energy gap at 0.3 GPa in Fig.~\ref{fig2}(b) becomes softened when approaching 
the ordered state~\cite{Hayashida2018}. 
Qualitatively different spectrum is observed in the ordered state at 1.4 GP in Fig.~\ref{fig2}(c).
A well-defined gapless excitation emerges at $k = 1/3$ and 2/3 
and another dispersive excitation with the minimum energy transfer ($\hbar \omega$) of 0.55 meV 
are observed in the higher energy range. 

\begin{figure}[tbp]
\begin{center}
\epsfig{file=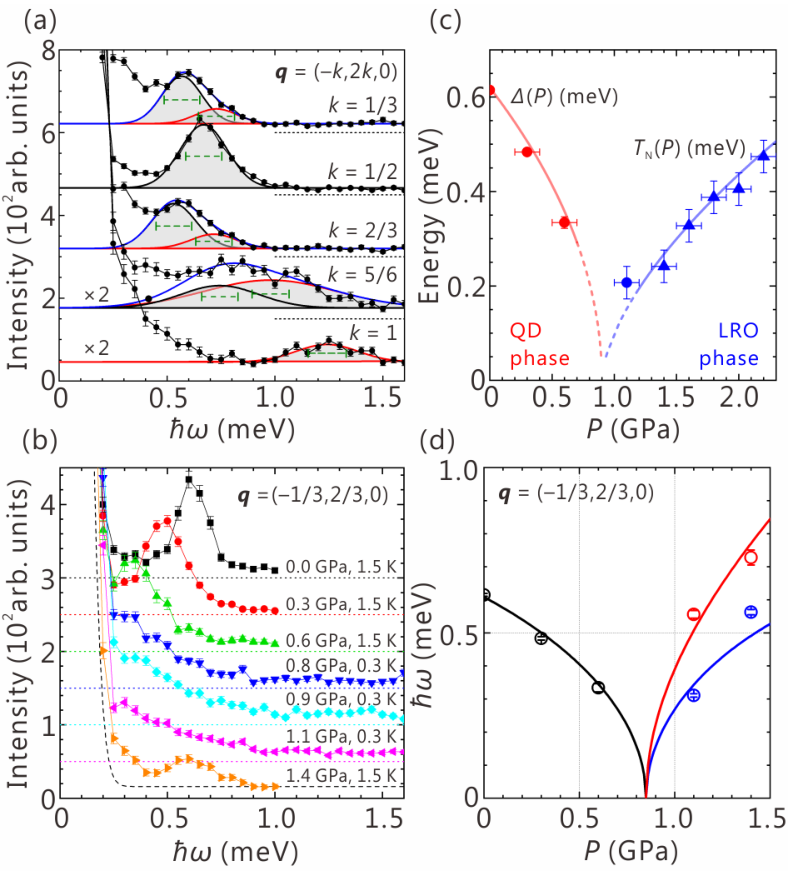,width=0.5\hsize}
\caption{(a) Constant-${\bm q}$ scans at $(-k,2k,0)$ under 1.4 GPa measured at a triple-axis spectrometer. 
Blue curves are the fitting result by Gaussian/pair Gaussian. 
Green dashed bars indicate the experimental resolution. 
(b) Pressure evolutions of the constant-${\bm q}$ scans at $(-1/3,2/3,0)$ 
obtained at a triple-axis spectrometer. The black dashed curve is a Gaussian function of the incoherent
scattering at 1.4 GPa with the FWHM of 0.17 meV.
(c) Phase diagram of the pressure-induced QPT in CsFeCl$_{3}$, which is obtained from 
energy gaps $\Delta$ and transition temperatures 
$T_{\rm N}$~\cite{Hayashida2018}. Red and blue curves are guides for eyes.
(d) Pressure dependence of the excitation energies at $(-1/3,2/3,0)$ calculated by the ESW.
Above 0.9 GPa, the blue and red curves are the excitations of gapless and gapped modes.
The circles are peak energies evaluated from the constant-${\bm q}$ scans.
}
\label{fig4}
\end{center}
\end{figure}

The INS spectra at 1.4 GPa were collected also by using a triple-axis spectrometer 
in order to cover wide $\hbar\omega$ - $\bm q$ range as shown in 
Fig.~\ref{fig4}(a). 
The spectral lineshapes at $k$ = 1/3 and 2/3 are rather asymmetric, 
and they are fitted by double Gaussians. 
Broad excitation is observed at $k$ = 5/6, and they can also be reproduced by double Gaussians, 
the widths of which are substantially wider than the experimental resolution. 
The peak energies are overplotted in Fig.~\ref{fig2}(c) by red diamonds. 
The excitation looks similar to the single dispersive excitations with the anisotropy gap at 0.0 and 0.3 GPa. 

The pressure evolution of the energy gap at ${\bm q} = (-1/3,~2/3,~0)$ is shown in
Fig.~\ref{fig4}(b).
The excitations at 0.0, 0.3, and 0.6 GPa show that the gap is suppressed with the increase 
of pressure.
At 0.8, 0.9, and 1.1 GPa, the gap position cannot be identified, and the broad excitations 
are observed below 1.0 meV. 
This implies that the system is near the QCP at these pressures.
At 1.4 GPa, the excitation appears at 0.55 meV, which corresponds to 
the high energy mode in Fig.~\ref{fig2}(c). 
The intensity of the excitation at 1.4 GPa in Fig.~\ref{fig4}(b) is weak compared with those in the 
QD phase.
This is because the most of the magnetic intensity is concentrated at the magnetic Bragg peak due to 
the long-range ordering.
The energies of the gap at 0.0, 0.3, and 0.6 GPa and the transition temperatures obtained both 
in the present and previous~\cite{Hayashida2018} studies are plotted in Fig.~\ref{fig4}(c).

\begin{figure}[t]
\begin{center}
\epsfig{file=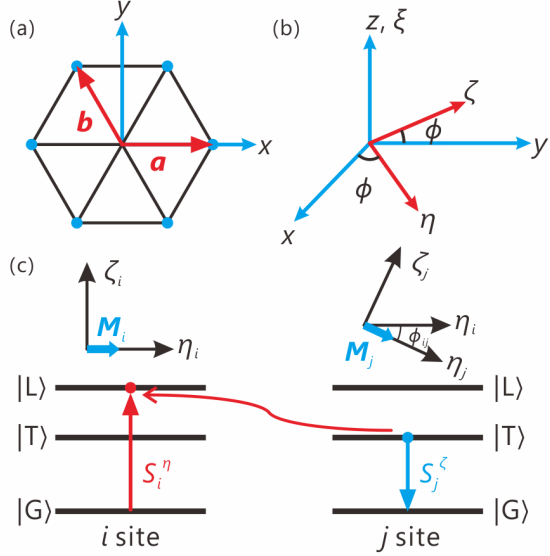,width=0.5\hsize}
\end{center}
\caption{
(a) Relation between the global $xyz$ coordinate and the crystallographic axes. 
(b) Relation between the global $xyz$ coordinate and the local $\eta \zeta \xi$ coordinate.
$\phi$ is the angle of the local magnetic moment measured from the $x$-axis.
$\eta$-axis ($\zeta$-axis) is taken parallel (perpendicular) to the moment in the $ab$ plane.
(c) Schematic of transition from $|T\rangle$ state to $|L\rangle$ states via the intersite interaction.
$\bm{M}_i$ represents the magnetic moment at the $i$ site.}
\label{fig:hopping}
\end{figure}

In order to discuss the obtained spectra, the one-magnon cross section was calculated 
based on the extended spin-wave theory (ESW)~\cite{Shiina2003}.
It is equivalent to the harmonic bond-operator theory~\cite{Sachdev1990,Sommer2001,Matsumoto2002,Matsumoto2004},
and is convenient to apply complex spin systems such as noncollinear ordered states~\cite{Matsumoto2014}.
For \Cs, we study the following Hamiltonian:
\begin{equation}
\H = \sum_i D \left( S_i^z \right)^2
+ J_c \sum_{\langle i,j\rangle}^{\rm chain} {\bm S}_i \cdot {\bm S}_j
+ J_{ab} \sum_{\langle i,j\rangle}^{\rm plane}  {\bm S}_i \cdot {\bm S}_j,
\label{eqn:H-s=1}
\end{equation}
where the sum is taken over the exchange interactions $J_{c}$ and $J_{ab}$ as shown in Fig.~\ref{fig1}(b). 
The relation between the crystallographic axes and the global $xyz$ coordinate of the spin system is shown in Fig.~\ref{fig:hopping}(a).
Here, we focus on the last term and rewrite it as
$\H_{ab}=\sum_{\langle i,j \rangle}^{\rm plane} \H_{ij}^{ab}$ with $\H_{ij}^{ab}=J_{ab}\bS_i\cdot\bS_j$.
Introducing creation and annihilation Bose operators
for the local longitudinal ($|L\rangle$) and transverse ($|T\rangle$) excited states
\cite{Shiina2003},
we can see that $\H_{ij}^{ab}$ brings about dynamics of the excited states as
$
\H_{ij}^{ab} = J_{ab}  \sum_{mn=L,T} \langle m | \bS_i |G\rangle \cdot \langle G| \bS_j | n\rangle a_{im}^\dagger a_{jn}.
$
In the local $\eta\zeta\xi$ coordinates shown in Fig. \ref{fig:hopping}(b), $\H_{ij}^{ab}$ is expressed as
\begin{align}
\H_{ij}^{ab} =J_{ab} \left[ \cos\phi_{ij} ( S_i^{\eta} S_j^{\eta} + S_i^{\zeta} S_j^{\zeta} ) + S_i^{\xi} S_j^{\xi}
+ \sin\phi_{ij} ( S_i^{\eta} S_j^{\zeta} - S_i^{\zeta} S_j^{\eta} ) \right] .
\label{eqn:H-XYZ-2}
\end{align}
Here, $\phi_{ij}=\phi_i-\phi_j$, and $\phi_i$ represents the angle of the magnetic moment at the $i$ site.
The first two ($\eta\eta+\zeta\zeta$ and $\xi\xi$) terms are diagonal for the $|L\rangle$ and $|T\rangle$ states,
while the last cross ($\eta\zeta-\zeta\eta$) term leads to hybridization between the $|L\rangle$ and $|T\rangle$ states
(LT-hybridization).
For instance, the $|T\rangle$ state can move from the $j$ site to the $i$ site and change into the $|L\rangle$ state
(see Fig. \ref{fig:hopping}(c)).
This process is described as
$
J_{ab} \sin\phi_{ij} \langle L | S_i^\eta |G\rangle \langle G| S_j^\zeta |T\rangle a_{iL}^\dagger a_{jT}.
$
In the same way, pair creation process is described as
$
J_{ab} \sin\phi_{ij} \langle L | S_i^\eta |G\rangle \langle T| S_j^\zeta |G\rangle a_{iL}^\dagger a_{jT}^\dagger.
$
The detailed description of the processes is summarized in Fig.~S2 in the supplementary material. 
We emphasize that one-magnon processes for the LT-hybridization
can only exist in noncollinear states, i.e. $\sin\phi_{ij}\neq 0$ in Eq. (\ref{eqn:H-XYZ-2}).

The spin interactions and anisotropy are parametrized by comparing the experiment and calculation, 
and they are represented as a function of the pressure by the linear-interpolation as follow: 
$J_{c}~({\rm meV})=-0.5-0.14 \times p,$ $J_{ab}~({\rm meV})=0.0312-0.0015 \times p,$ 
and $D~({\rm meV})=2.345 + 0.365 \times p$, 
where the $p$ (GPa) is the value of pressure.
Pressure dependences of the excitation energies at ${\bm q} = (-1/3,~2/3,~0)$ 
are indicated by circles and solid curves in Fig.~\ref{fig4}(d).
The data are reasonably reproduced by the calculation within the linear pressure dependence in $J_c$, $J_{ab}$, and $D$. 
The calculated dispersion relations obtained by using the extracted parameters are indicated by the solid yellow curves in Fig.~\ref{fig2}(a)-\ref{fig2}(c), and the calculation is consistent with the experiment both in the QD and LRO phases. 
Furthermore, the calculated INS spectra in Figs. \ref{fig2}(d)-\ref{fig2}(f) also reproduce the observed ones. 

To understand effects of the LT-hybridization, we demonstrate the INS spectra after dropping the cross term in Eq.~(\ref{eqn:H-XYZ-2}).
The results are shown in Figs. \ref{fig2}(h) and \ref{fig2}(i),
where the gapless (gapped) modes are pure transverse (longitudinal) modes in this case.
With the increase of pressure from 1.4 to 4.0 GPa, the longitudinal modes shift to high-energy region and lose intensity,
as in the collinear case of \Tl
\cite{Ruegg2008}.
When the LT-hybridization is taken into account,
off-diagonal elements between the $|L\rangle$ and $|T\rangle$ states lead to level repulsion.
Far from the QCP ($p$ = 4.0 GPa), the INS spectrum is not affected by the hybridization
[compare Figs. \ref{fig2}(g) and \ref{fig2}(h)].
By contrast, the spectrum is strongly renormalized by the hybridization near the QCP 
[compare Figs. \ref{fig2}(f) and \ref{fig2}(i)] and novel magnetic excitations are formed; the both gapless and gapped
modes are accompanied by strong longitudinal and transverse fluctuations. 
Interestingly, shape of the INS spectrum resembles that in the QD phase [compare Figs. \ref{fig2}(e) and \ref{fig2}(f)].
Note that the spectrum continuously evolves through the QCP and the property of the second-order phase transition is ensured
by taking the LT-hybridization into account.
Thus, the LT-hybridization plays an important role in magnon dynamics in noncollinear magnets near QCP.
The INS measurements in \Cs~revealed this by the fine tuning of pressure through the QCP.

Since the newly found excitation exists in a noncollinear spin structure, the search of the excitation in different types of noncollinear structures such as cycloidal structure, all-in all-out structure, skyrmion lattice, etc., would be interesting topics. 
Search of the hybridized mode in other systems including charge density wave, spin density wave, ultra-cold atoms, etc., would be  important. 
Finally, the effect of the hybridization to the lifetime of the magnon and other elementary excitations would be also interesting. 



\section*{Acknowledgements}
We are grateful to D. Kawana, T. Asami, R. Sugiura, A. A. Aczel, S. Asai, and S. Hasegawa for supporting the neutron scattering experiments.
Prof. M. Takigawa is greately appreciated for fruitful discussions. 
The neutron scattering experiment at the HRC was approved by the Neutron Scattering
Program Advisory Committee of IMSS, KEK (Proposals No.~2015S01, No.~2016S01 and No.~2017S01) and ISSP.
The neutron scattering experiment at the CTAX used resources at the High Flux Isotope Reactor, a DOE office of Science
User Facility operated by the ORNL (IPTS-16770.1). 
Travel expenses for the neutron scattering experiments performed using CTAX at
ORNL, USA was supported by the U.S.-Japan Cooperative Research Program on Neutron Scattering (Proposal No.~2017-21).
S.H. was supported by the Japan Society for the Promotion of  Science through the 
Leading Graduate Schools (MERIT).
M.M. was supported by JSPS KAKENHI Grant Number 17K05516, and 
N.K. and H.T. were supported by JSPS KAKENHI Grant Number 16K05414.

\section*{Supplementary materials}
Materials and Methods\\
Figs. S1 to S3\\
Table S1\\
References \textit{(28-33)}


\end{document}



\baselineskip24pt


\maketitle 

\setcounter{equation}{0}
\setcounter{figure}{0}
\setcounter{table}{0}
\setcounter{page}{1}
\makeatletter
\renewcommand{\theequation}{S\arabic{equation}}
\renewcommand{\thefigure}{S\arabic{figure}}
\renewcommand{\thetable}{S\arabic{table}}
\renewcommand{\H}{{\mathcal H}}

\newcommand{\bS}{{\bm{S}}}
\newcommand{\bW}{{\bm{W}}}

\newcommand{\bk}{{\bm{k}}}
\newcommand{\br}{{\bm{r}}}
\newcommand{\bd}{{\bm{d}}}
\newcommand{\bq}{{\bm{q}}}
\newcommand{\bQ}{{\bm{Q}}}

\newcommand{\ba}{{\bm{a}}}
\newcommand{\bb}{{\bm{b}}}
\newcommand{\bc}{{\bm{c}}}
\newcommand{\bA}{{\bm{A}}}
\newcommand{\bB}{{\bm{B}}}
\newcommand{\bu}{{\bm{u}}}
\newcommand{\bv}{{\bm{v}}}
\newcommand{\balpha}{{\bm{\alpha}}}
\newcommand{\bU}{{\bm{U}}}
\newcommand{\bV}{{\bm{V}}}

\newcommand{\hq}{{\hat{q}}}

\newcommand{\TN}{T_{\rm N}}

\newcommand{\Sr}{{Sr$_2$CoSi$_2$O$_7$}}
\newcommand{\Ba}{{Ba$_2$CoGe$_2$O$_7$}}
\newcommand{\Tl}{{TlCuCl$_3$}}
\newcommand{\Cs}{{CsFeCl$_3$}}
\newcommand{\Rb}{{RbFeCl$_3$}}
\newcommand{\BaCo}{{Ba$_3$CoSb$_2$O$_9$}}


{\bf This PDF file includes: }
\\
Materials and Methods\\
Figs. S1 to S3\\
Table S1 \\
References

%
%
%


\section*{Materials and Methods}

\subsection*{Experimental method}
A single crystal sample was grown by the vertical Bridgman method~({\it 21}).
The sample was mounted in a Teflon cell with a deuterated glycerin as a pressure 
medium so that the horizontal plane is the $ab$ plane. 
The Teflon cell was installed in a piston-cylinder clamped cell made of CuBe-alloy.
We performed the INS study under pressures using 
the high resolution chopper spectrometer (HRC) at J-PARC/MLF in Japan~\cite{Itoh2011,Yano2011,Itoh2013} 
and cold neutron triple-axis spectrometer CTAX at ORNL in USA. 
The details of the experimental setups including the name of the spectrometer, sample mass, applied pressure, the energy of the neutron, and the temperature are summarized in Table~\ref{tb1}. 
\begin{table}[htbp]
\begin{tabular}{cccccc}
\hline 
\hline
Setup & Instrument & Sample mass (g) & Pressure (GPa) &  $E_{i}$ or $E_{f}$ (meV) & Base Temperature (K)  \\
\hline
1 & HRC & 0.75 & 0.0 &  8.11 &  6 \\
2 & HRC & 0.35 & 0.3 & 5.08 &  2.7 \\
3 & HRC &  0.46 & 1.4 & 3.05 &  0.9 \\
4 & CTAX &  0.46 & 1.4 & 3.5 &  1.5 \\
5 & CTAX & 0.35 & 0.0, 1.4 & 3.5 &  1.5 \\
6 & CTAX & 0.35 & 0.3 & 3.5 & 1.5  \\
7 & CTAX & 0.37  & 0.6 & 3.5 & 1.5  \\
8 & CTAX & 0.37 & 0.8, 0.9, 1.1 & 3.5 & 0.3 \\
 \hline \hline
\end{tabular} 
\caption{The different setups used for INS measurements under pressures.
INS measurements were operated with fixing energies of incident neutron $E_{i}$ for HRC
and final neutron $E_{f}$ for CTAX.
Regarding to cryostats, we used the closed cycle refrigerator for Setups 1 and 2, the $^{4}$He
cryostat for Setups 4, 5, 6 and 7, and $^{3}$He cryostat for Setups 3 and 8.}
\label{tb1}
\end{table}

\begin{figure*}[htbp]
\epsfig{file=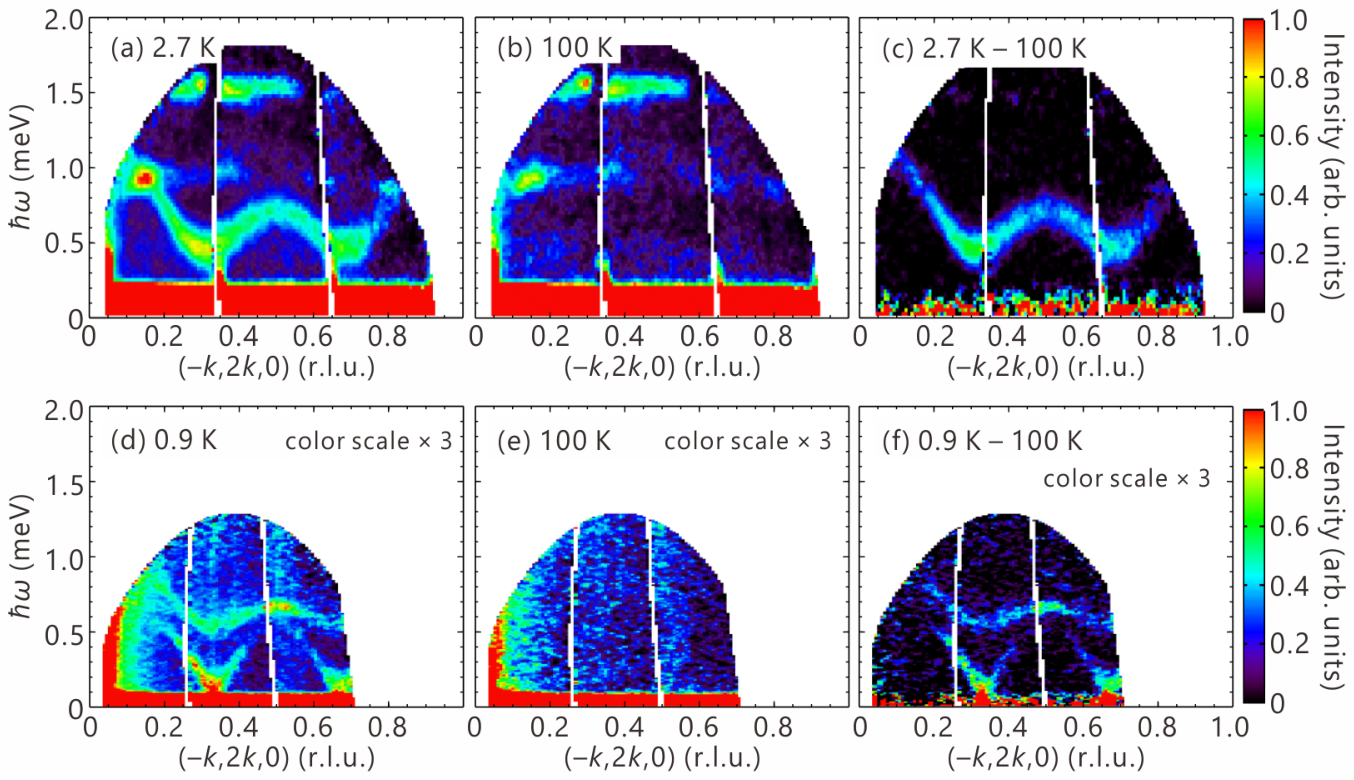,width=1.0\hsize}
\caption{Temperature evolution of the INS spectra obtained at HRC at (a) 2.7 K, 
(b) 100 K under 0.3 GPa and at (d) 0.9 K and (e) 100 K under 1.4 GPa.
The INS spectra under (c) 0.3 GPa and (h) 1.4 GPa are subtracted by the backgrounds
at 100K.}
\label{suppl1}
\end{figure*}

In the HRC experiment, the pressure cell was installed into the closed-cycle $^4$He cryostat for 0.0 and
0.3 GPa, and the $^{3}$He cryostat for 1.4 GPa.
The energies of the incident neutron were $E_{i}=8.11$ meV for 0.0 GPa,  
$E_{i}=5.08$ meV for 0.3, and $E_{i}=3.05$ meV for 1.4 GPa.
Figures~\ref{suppl1}(a) and \ref{suppl1}(b) show the INS spectra at 2.7 K and 100 K
under 0.3 GPa measured at HRC.
At 2.7 K, flat excitations are observed at 0.9 meV and 1.5 meV in addition to the dispersive
excitation below 1.0 meV.
These flat excitations remain at 100 K even though the dispersive one disappears.
The INS spectra at 0.9 K and 100 K under 1.4 GPa are shown in Figs.~\ref{suppl1}(d) and
\ref{suppl1}(e).
At 0.9 K, strong intensity is observed around ${\bm q}=0$ in addition to the dispersive
excitation, and it remains at 100 K.
We thus presume that the data measured at 100 K was the background from the pressure cell and cryostat.
Then, the data at base temperatures for 0.3 GPa and 1.4 GPa were subtracted by the
background as shown in Figs.~\ref{suppl1}(c) and \ref{suppl1}(f).

In the CTAX experiment, the cell was installed into the $^{4}$He cryostat for 0.0, 0.3, 0.6, and 1.4 GPa,
and $^{3}$He cryostat for 0.8, 0.9, and 1.1 GPa.
The energy of the final neutron was fixed to be $E_{f}=3.5$ meV.

\subsection*{Extended Spin-Wave Theory for Triangular Antiferromagnet}

\subsubsection*{Hamiltonain}

The spin Hamiltonian for the triangular antiferromagnet \Cs~is given by the following form:
\begin{align}
&\H = \sum_i \H^D_i
+ \sum_{\langle i,j\rangle}^{\rm chain} \H^c_{ij}
+ \sum_{\langle i,j\rangle}^{\rm plane} \H^{ab}_{ij},
\label{eqn:H0}
\end{align}
where
\begin{align}
&\H^D_i = D \left( S_i^z \right)^2,~~~
\H^c_{ij} = J_c \bS_i\cdot\bS_j,~~~
\H^{ab}_{ij} = J_{ab} \bS_i\cdot\bS_j.
\label{eqn:H-s=1}
\end{align}
In Eq. (\ref{eqn:H0}), the summation $\sum_{\langle i,j\rangle}^{\rm chain}$ and $\sum_{\langle i,j\rangle}^{\rm plane}$ are taken
over the nearest neighbor spin pairs along the $c$-axis and in the $ab$-plane, respectively.
$J_c$ is strong ferromagnetic ($J_c<0$) and $J_{ab}$ is weak antiferromagnetic ($J_{ab}>0$).

As observed by the neutron scattering measurement, the 120$^\circ$ structure in the $ab$-plane is stabilized.
In this structure, the angle of the magnetic moment measured from the $x$-axis is expressed as
$\phi_i = \bQ\cdot\br_i + \phi_0$ with $\bQ = \frac{1}{3} \ba^* +\frac{1}{3} \bb^*$.
Here, $\br_i$ represents the position of the $i$ site.
$\ba^*$ and $\bb^*$ are reciprocal lattice vectors.
$\phi_0$ represents the phase of the 120$^\circ$ structure.
It does not affect the result.

For the 120$^\circ$ strucure, we introduce a local $\eta \zeta \xi$ coordinate at each $i$ site.
Here, the $\eta$-axis is taken along the ordered moment on the $xy$-plane, as shown in Fig. 4(b) 
in the original paper~({\it 18,32,33}).
The $\zeta$-axis is chosen perpendicular to the $\eta$-axis on the $xy$-plane.
In the $\eta \zeta \xi$ coordinate, the spin operator is transformed as
\begin{align}
\begin{pmatrix}
S_i^x \cr
S_i^y \cr
S_i^z
\end{pmatrix}
=
\begin{pmatrix}
\cos\phi_i & -\sin\phi_i & 0 \cr
\sin\phi_i & \cos\phi_i & 0 \cr
0 & 0 & 1
\end{pmatrix}
\begin{pmatrix}
S_i^{\eta} \cr
S_i^{\zeta} \cr
S_i^{\xi}
\end{pmatrix}.
\label{eqn:S-trans}
\end{align}
The Hamiltonian is then expressed as
\begin{align}
&\H^D_i = D \left( S_i^\xi \right)^2, \cr
&\H^c_{ij} = J_c \left( S_i^{\eta} S_j^{\eta} + S_i^{\zeta} S_j^{\zeta} + S_i^{\xi} S_j^{\xi} \right), \cr
&\H^{ab}_{ij} = J_{ab} \cos{\phi_{ij}} ( S_i^{\eta} S_j^{\eta} + S_i^{\zeta} S_j^{\zeta} + S_i^{\xi} S_j^{\xi} )
+ J_{ab} \sin{\phi_{ij}} ( S_i^{\eta} S_j^{\zeta} - S_i^{\zeta} S_j^{\eta} ).
\label{eqn:H-XYZ-2}
\end{align}
Here, $\phi_{ij}=\phi_i-\phi_j$.
The longitudinal and transverse spin fluctuations of the ordered moment are decoupled
in the diagonal $S_i^{\eta} S_j^{\eta} + S_i^{\zeta} S_j^{\zeta} + S_i^{\xi} S_j^{\xi}$ term.
On the other hand, they are hybridized by the cross $S_i^{\eta} S_j^{\zeta} - S_i^{\zeta} S_j^{\eta}$ term
in noncollinear states $(\sin\phi_{ij}\neq 0)$, as we will see below.

\subsubsection*{Mean-field solution}

At each site, the mean-field Hamiltonian is expressed as
\begin{align}
\H^{\rm MF} = D \left( S^{\xi} \right)^2 + \left( 2J_c + 3J_{ab} \right) S^{\eta} \langle S^{\eta} \rangle.
\label{eqn:H-mf}
\end{align}
Here, $\langle S^{\eta} \rangle$ represents the expectation value of the moment of the local ground state.
Note that the mean-field solution is the same at all site in the local $\eta\zeta\xi$ coordinate.
The mean-field ground and excited states are given by ({\it 18})
\begin{align}
&|G\rangle = u |0\rangle + \frac{v}{\sqrt{2}} \left( |1\rangle + |-1\rangle \right), \cr
&|L\rangle = - v |0\rangle + \frac{u}{\sqrt{2}} \left( |1\rangle + |-1\rangle \right), \cr
&|T\rangle = \frac{1}{\sqrt{2}} \left( - |1\rangle + |-1\rangle \right).
\end{align}
Here, $|G\rangle$ is the ground state,
while the longitudinal $|L\rangle$ and the transverse $|T\rangle$ states are the second and the first excited states, respectively.
The expectation value of the moment is expressed as $\langle S^{\eta} \rangle = \langle G | S^{\eta} | G \rangle$.
In the disordered phase, $u=1$ and $v=0$.
In the ordered phase, they are given by ({\it 18})
\begin{align}
u = \sqrt{ \frac{1}{2} \left( 1 + \frac{D}{J_{\rm eff}} \right) },~~~
v = \sqrt{ \frac{1}{2} \left( 1 - \frac{D}{J_{\rm eff}} \right) },
\end{align}
with
\begin{align}
J_{\rm eff} = 2 ( 2 J_c + 3 J_{ab} ).
\end{align}
The parameters $D$, $J_c$, and $J_{ab}$ change with pressure.
The quantum critical point is expressed by $D = J_{\rm eff}$.

On the basis of the $|G\rangle$, $|L\rangle$, and $|T\rangle$ states,
spin operators at each site are expressed in the following matrix form ({\it 18}):
\begin{align}
S^\eta =
\begin{pmatrix}
2uv     & u^2-v^2 & 0\cr
u^2-v^2 & -2uv & 0 \cr
0 & 0 & 0
\end{pmatrix},~~~
S^\zeta =
\begin{pmatrix}
0  & 0 & -iu \cr
0 & 0   & iv \cr
iu  & -iv  & 0
\end{pmatrix},~~~
S^\xi =
\begin{pmatrix}
0  & 0  & -v \cr
0  & 0  & -u \cr
-v & -u & 0
\end{pmatrix}.
\end{align}
As shown in Fig. 1(a) in the original paper,
the $|L\rangle$ state can be excited from the ground state by the longitudinal $(S^\eta)$ component,
while the $|T\rangle$ state is excited by the transverse  ($S^\zeta$ and $S^\xi$) components.
This indicates that the $|L\rangle$ and $|T\rangle$ states
are accompanied by the longitudinal and transverse spin fluctuations, respectively.
Since $u\gg v$ in the vicinity of the quantum critical point, $S^\zeta$ dominates the excitation of the $|T\rangle$ state.
This means that the transverse fluctuation is restricted in the $ab$ plane by the strong easy-plane anisotropy.

\subsubsection*{Magnetic excitations}

The extended spin-wave theory (ESW) was introduced to study excited states in quadrupole ordered phase in $f$ 
electrons systems ({\it 23, 34}). 
It is equivalent to the harmonic bond-operator formulation ({\it 24})
introduced to study bilayer spin 
system ({\it 25}), 
which was also applied to interacting spin dimer systems ({\it 26, 27}). 
The ESW has advantage to study spin systems with various types of complex anisotropies
\cite{Matsumoto-2008}.
We briefly introduce the formulation and explain
how the hybridization of the $|L\rangle$ and $|T\rangle$ states is brought about in one-magnon process in noncollinear states.

The characteristic point of the ESW is that bosons are introduced for the local excited states ({\it 23}).
On the basis of the mean-field solution, the mean-field Hamiltonian at each site is expressed as ({\it 18})
\begin{align}
\H^{\rm MF}_i
= \sum_{m=T,L} E_m a_{im}^\dagger a_{im},
\label{eqn:H-local}
\end{align}
where
$E_m$ is the energy eigenvalue of the excited state measured from the ground state.
$a_{im}^\dagger$ and $a_{im}$ are creation and annihilation Bose operators for the $|m\rangle$ state
at the $i$ site, respectively ({\it 23}).
Equation (\ref{eqn:H-local}) represents that there are two ($m=L,T$) local excited states at each site.

\begin{figure}[t]
\begin{center}
\includegraphics[width=13cm,clip]{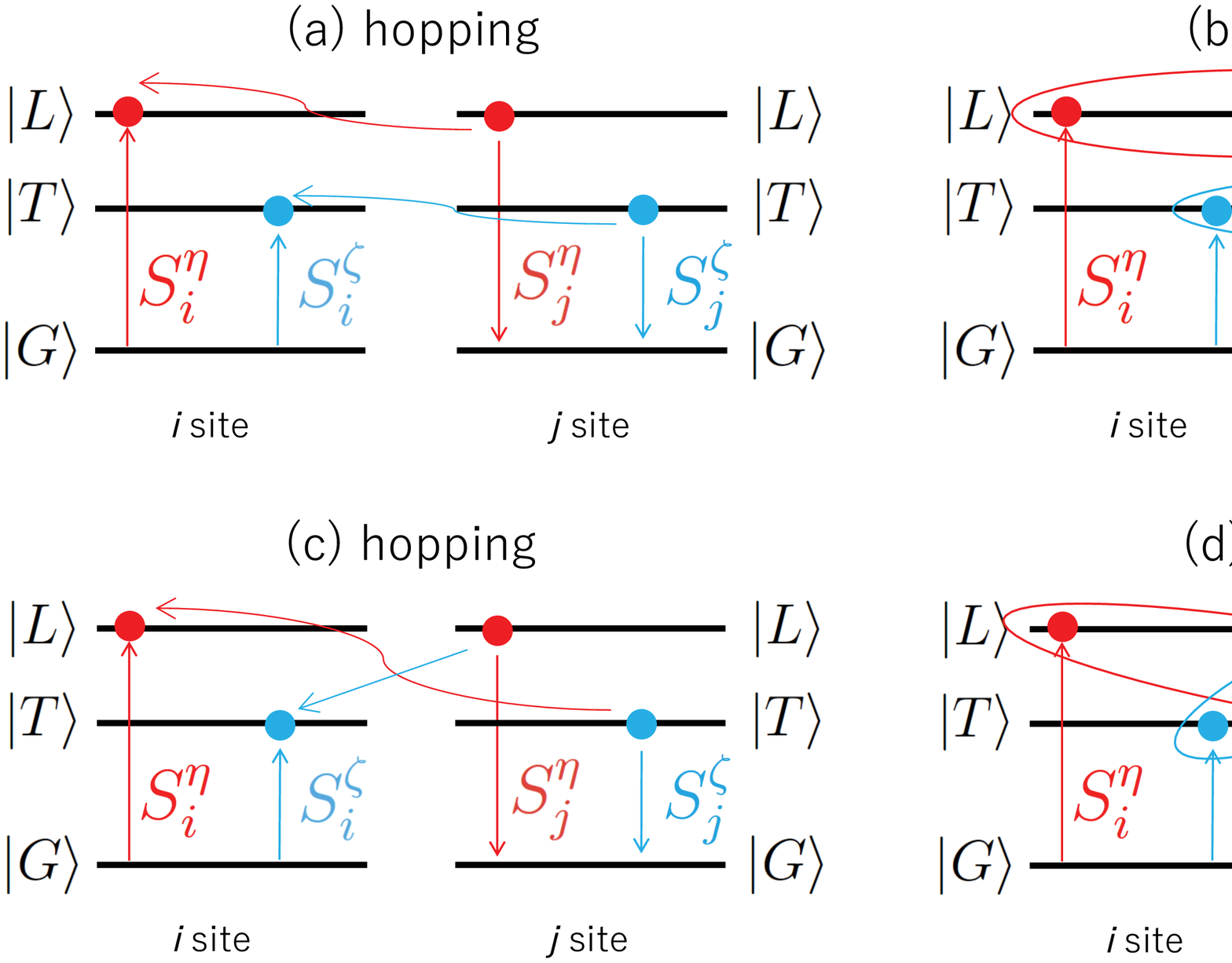}
\end{center}
\caption{
Schematic of hopping and pair creation processes.
(a) and (b) are for diagonal processes for the $|L\rangle$ and $|T\rangle$ states.
(c) and (d) are for off-diagonal processes for the $|L\rangle$ and $|T\rangle$ states.
The ellipse represents the created pair of excited states.
}
\label{fig:process}
\end{figure}

The intersite interactions give rise to dynamics of bosons.
First, we consider $\H^c_{ij}$ in Eq. (\ref{eqn:H-XYZ-2}).
The hopping process is expressed as
\begin{align}
\H^{c({\rm hop})}_{ij} &= J_c \sum_{\alpha=\eta,\zeta,\xi} \sum_{m=T,L}
|m\rangle \langle m | S_i^\alpha |G\rangle \langle G| S_j^\alpha \sum_{n=T,L} |n\rangle \langle n| + {\rm h.c.} \cr
&= J_c \sum_{mn=T,L} \sum_{\alpha=\eta,\zeta,\xi}
\langle m | S_i^\alpha |G\rangle \langle G| S_j^\alpha |n\rangle a_{im}^\dagger a_{jn} + {\rm h.c.} \cr
&= J_c \left[ \langle L | S_i^\eta |G\rangle \langle G| S_j^\eta |L\rangle a_{iL}^\dagger a_{jL}
+ \sum_{\alpha=\zeta,\xi} \langle T | S_i^\alpha |G\rangle \langle G| S_j^\alpha |T\rangle a_{iT}^\dagger a_{jT} \right] + {\rm h.c.}
\label{eqn:hop}
\end{align}
Here, the first (second) term describes hopping of the $|L\rangle$ ($|T\rangle$) state.
It is caused by the longitudinal (transverse) component of the spin operator, as shown in Fig. \ref{fig:process}(a).
As for the transverse component, $S^\zeta$ dominates the excitation.
In addition to the hopping process, there are pair creation and annihilation processes.
They are expressed as
\begin{align}
\H^{c({\rm pair})}_{ij}
&= J_c \sum_{mn=T,L} \sum_{\alpha=\eta,\zeta,\xi}
\langle m | S_i^\alpha |G\rangle \langle n| S_j^\alpha |G\rangle a_{im}^\dagger a_{jn}^\dagger + {\rm h.c.} \cr
&= J_c \left[ \langle L | S_i^\eta |G\rangle \langle L| S_j^\eta |G\rangle a_{iL}^\dagger a_{jL}^\dagger
+ \sum_{\alpha=\zeta,\xi} \langle T | S_i^\alpha |G\rangle \langle T| S_j^\alpha |G\rangle
a_{iT}^\dagger a_{jT}^\dagger \right] + {\rm h.c.}
\label{eqn:pair}
\end{align}
These processes are shown in Fig. \ref{fig:process}(b).
Since the cosine term of $\H^{ab}_{ij}$ in Eq. (\ref{eqn:H-XYZ-2}) is diagonal for the $|L\rangle$ and $|T\rangle$ states,
the hopping, pair creation, and pair annihilation processes are also expressed by Eqs. (\ref{eqn:hop}) and (\ref{eqn:pair})
with the replacement of $J_c\rightarrow J_{ab} \cos\phi_{ij}$.
Therefore, the $|L\rangle$ and $|T\rangle$ states are not hybridized by the above processes.
On the other hand, the two states are hybridized by the sine term of $\H^{ab}_{ij}$ as in the following from:
\begin{align}
\H^{ab(\sin)({\rm hop})}_{ij} &= J_{ab} \sin{\phi_{ij}} \left[
\langle L | S_i^\eta |G\rangle \langle G| S_j^\zeta |T\rangle a_{iL}^\dagger a_{jT}
- \langle T | S_i^\zeta |G\rangle \langle G| S_j^\eta |L\rangle a_{iT}^\dagger a_{jL} \right] + {\rm h.c.}, \cr
\H^{ab(\sin)({\rm pair})}_{ij} &= J_{ab} \sin{\phi_{ij}} \left[
\langle L | S_i^\eta |G\rangle \langle T| S_j^\zeta |G\rangle a_{iL}^\dagger a_{jT}^\dagger
- \langle T | S_i^\zeta |G\rangle \langle L| S_j^\eta |G\rangle a_{iT}^\dagger a_{jL}^\dagger \right] + {\rm h.c.}
\label{eqn:LT}
\end{align}
These processes are shown in Figs. \ref{fig:process}(c) and \ref{fig:process}(d).
We emphasize that such processes for the LT-hybridization can only exist
in noncollinear ordered states, i.e. $\sin\phi_{ij}\neq 0$,
and play important role in magnon dynamics in the vicinity of the quantum critical point.

Dispersion relations of the magnetic excitation modes,
dynamical spin correlation function, and INS spectra are obtained by the ESW.
Details are described in Refs. 18, 23, and 35.

\subsubsection*{Evolution of Calculated Neutron Spectra by Pressure}

\begin{figure}[htbp]
\includegraphics[width=16cm,clip]{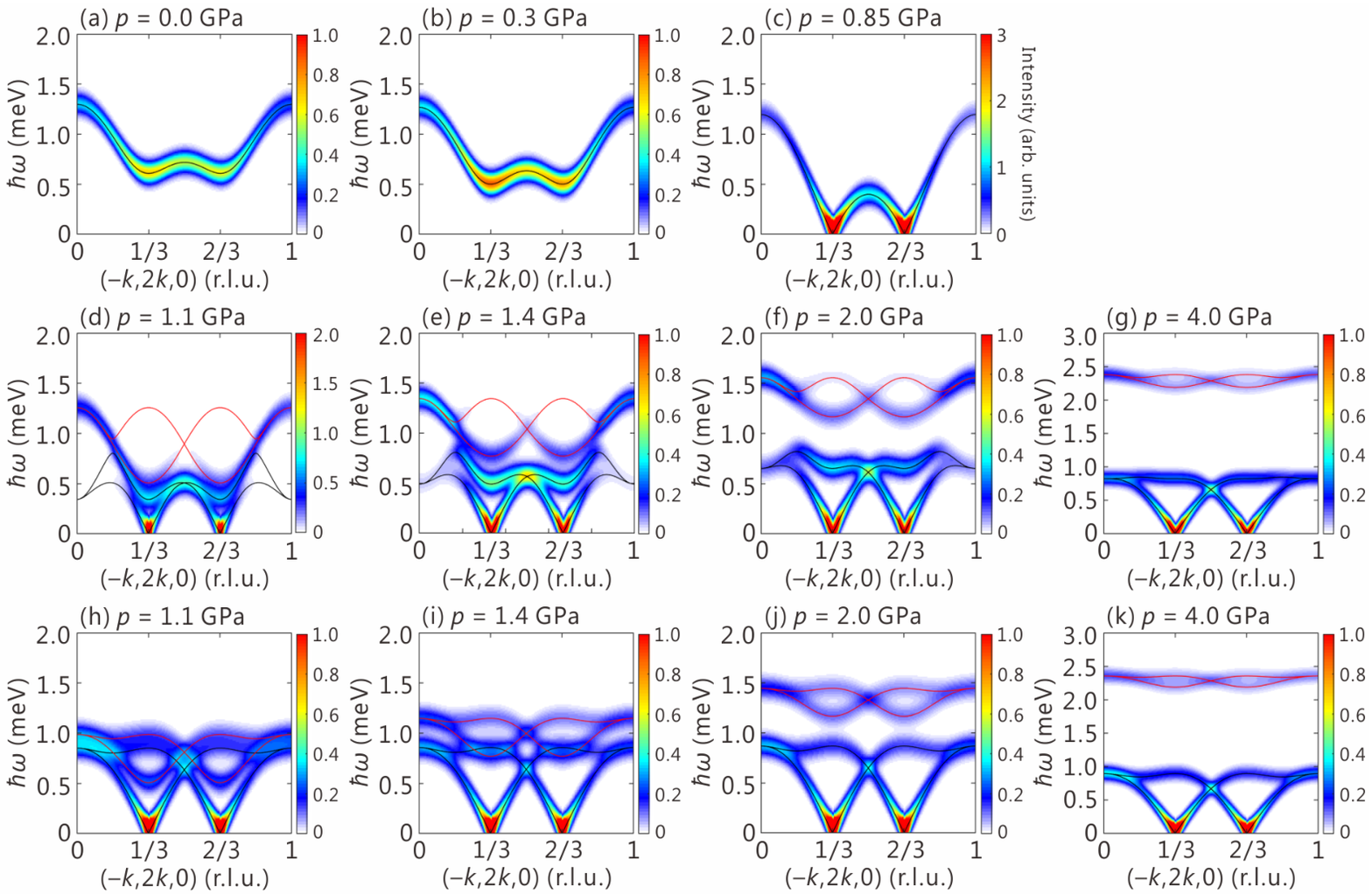}
\caption{
Calculated INS spectra at various pressures.
(a) for 0 GPa, (b) for 0.3 GPa, (c) for 0.85 GPa,  (d) for 1.1 GPa, (e) for 1.4 GPa, (f) for 2.0, and (g) for 4.0 GPa.
The critical pressure is around 0.85 GPa.
In the ordered phase, the red and black lines are
for the low-energy $E_{\bq\pm\bQ,1}$ and the high-energy $E_{\bq\pm\bQ,2}$ modes, respectively.
Calculated spectra dropping the LT-hybridization terms are also shown in 
(h) for 1.1 GPa, (i) for 1.4 GPa, (j) for 2.0 GPa, and (k) for 4.0 GPa.
}
\label{varP}
\end{figure}

As in the original paper, we introduce the following linear pressure dependences in $D$, $J_c$, and $J_{ab}$:
\begin{align}
D~({\rm meV}) = &2.345 + 0.365 \times p, \\
J_{c}~({\rm meV}) = &-0.5-0.14 \times p, \\
J_{ab}~({\rm meV}) = &0.0312-0.0015 \times p.
\end{align}
Calculated INS spectra at various pressure are shown in Figs.~\ref{varP}(a)-\ref{varP}(g).
In the disordered phase $(p<p_c)$, single dispersion having two-fold degeneracy is reproduced
with energy gaps at $\bq = (-1/3, 2/3, 0)$ and $(-2/3, 4/3, 0)$.
The gap is softened at the critical pressure $p_c(=0.85$ GPa).
In the ordered phase ($p > p_c$), the mode splits into low- and high-energy modes of $\omega=E_{\bq,1}$ and $E_{\bq,2}$, respectively.
The former is gapless, while the latter is gapped.

In noncollinear ordered states, $E_{\bq,\ell}$, $E_{\bq+\bQ,\ell}$, and $E_{\bq-\bQ,\ell}$ $(\ell=1,2$) modes
are observable in general by INS measurements.
In the \Cs~case, the $E_{\bq \ell}$ mode is observable via the dynamical spin correlation function $S^{zz}(\bq,\omega)$,
while the both $E_{\bq\pm\bQ \ell}$ modes are observable via $S^{xx}(\bq,\omega)$ and $S^{yy}(\bq,\omega)$.
Owing to the strong easy-plane anisotropy of the $D$ term,
the intensity of $S^{zz}(\bq,\omega)$ is strongly suppressed and the $E_{\bq,\ell}$ mode is hard to detect in \Cs.
This means that the spin fluctuations are restricted in the $ab$ plane
and we only plot the four $E_{\bq\pm\bQ,\ell}$ ($\ell$=1,2) modes in Fig. \ref{varP}.

Just above $p_c$, we show such split modes of $E_{\bq\pm\bQ \ell}$ ($\ell$=1,2) in Fig.~\ref{varP}(d).
We can notice that the strong intensity region traces the single mode in the disordered phase.
This indicates that the INS spectra continuously evolves through the critical pressure
and satisfies the property of the second-order phase transition.
At 1.4 GPa [see Fig.~\ref{varP}(e)], the low- and high-energy modes split more and the shape of the strong intensity
begins to deviate from the single mode in the disordered phase.
Under high pressures [see Figs.~\ref{varP}(f) and \ref{varP}(g)],
the gapped modes shift to higher energy and lose intensity.
At 1.4 GPa, where the present measurements were performed,
we can observe the intermediate feature of such continuous transition of the magnetic excitation
from the disordered to the ordered phases.
Thus, we notice that the observed high energy mode in Fig. 2(c) at 1.4 GPa in the original paper
is the remnant of the single mode in the disordered phase.

In the vicinity of the critical pressure, the energy levels of the $|L\rangle$ and $|T\rangle$ states are very close.
In addition to this, the matrix element for the LT-hybridization is large.
These lead to the strong hybridization between the $|L\rangle$ and $|T\rangle$ states in one-magnon level
and the excitation energies are strongly renormalized.
As the result, the novel excitations are formed, i.e. the both low- and high-energy modes are accompanied
by the strong longitudinal and transverse fluctuations of the ordered moment.
We can see this point by comparing Figs. \ref{varP}(d)-\ref{varP}(g) with Figs. \ref{varP}(h)-\ref{varP}(k),
where the LT-hybridization terms given by Eq. (\ref{eqn:LT}) are dropped in the latter figures.
When the hybridization terms are dropped, the formulation is not valid any more
and we fail in reproducing the continuous evolution of the INS spectra through the critical pressure.
Thus, the LT-hybridization cannot be ignored and is inevitable for the correct formulation.
It ensures the continuous evolution of the INS spectra
and leads to the novel excitation modes in the vicinity of the quantum critical point in noncollinear magnets.
